\documentclass[11pt,twoside,A4]{article} 
\usepackage{times,fancyhdr}
\usepackage[dvips]{graphicx}
\usepackage{latexsym} 
\usepackage[affil-it]{authblk}
\usepackage{amsmath}
\usepackage{amssymb}
\usepackage{setspace}

\usepackage[top=4cm, bottom=4cm, left=4cm, right=4cm]{geometry}

\def\beq{\begin{equation}}
\def\eeq{\end{equation}}
\def\beqn{\begin{eqnarray}}
\def\eeqn{\end{eqnarray}}

\begin{document}
 
\title{Disentangling the Black Hole Vacuum}
\author{Sabine Hossenfelder\thanks{hossi@nordita.org}} 
\affil{\small Nordita\\
KTH Royal Institute of Technology and Stockholm University\\
Roslagstullsbacken 23, SE-106 91 Stockholm, Sweden}
\date{}
\maketitle
\begin{abstract}
We study the question whether disentanglement of 
Hawking radiation can be achieved with any local operation. We assume
that the operation we look for is unitary and can be described by a
Bogoliubov transformation. This allows to formulate requirements
on the operation of disentanglement. We then show that these requirements
can be fulfilled by a timelike boundary condition in the near-horizon
area and that the local observer does not notice the presence of the
boundary and does not encounter a firewall. 
\end{abstract}

\section{Introduction}

The black hole information loss problem, ignited by Hawking's seminal work \cite{Hawking:1974sw,Hawking:1976ra}, has kept theoretical physicists busy for more than three decades
now. A new spin was put on this problem in a recent paper \cite{Almheiri:2012rt} (hereafter referred
to as {\sc AMPS}) that drew attention to an apparent inconsistency in one of the most popular solution
attempts to the black hole information loss problem, black hole complementarity \cite{Susskind:1993if}. Black
hole complementarity attempts to recover information that fell into the black hole by allowing
it to both fall in and remain available outside. {\sc AMPS} argued, using very few assumptions,
that a local observer at the horizon must notice if the radiation outgoing to infinity is not a
thermal mixed state. If black hole complementarity was correct, so the argument, the local
observer would encounter a `firewall', implying an inconsistency with the equivalence
principle. This inconsistency challenges our present understanding of black hole
physics and consequently has drawn a lot of attention. For a summary and
discussion of many follow-up works please refer to \cite{Almheiri:2013hfa}.

As previously pointed out in \cite{Hossenfelder:2012mr}, the central assumption used by {\sc AMPS} is that
the entanglement necessary for the purity of the outgoing radiation is of a specific
form and in particular contains correlations over long times. This assumption is reasonable in
that this is a state one `typically' expects, but raises the question of whether it is the
correct way to purify the mixed state.

In \cite{Verlinde:2013uja} it was suggested to resolve the black hole firewall puzzle by 
disentangling the Hawking modes inside and outside the horizon which realizes a specific purification procedure. The
disentanglement was proposed to be achieved
by the operation of a {\sc CNOT} swap that mixes annihilation and creation operators in a suitable way. This
swap is assumed to affect only the local observer. The
result is that the state at ${\cal I}^+$ can be pure even though the infalling observer does not notice any
deviation from the normally mixed radiation. 

However, besides the question of how, physically, such a swap would happen for the local observer only and how to
interpret it, it also
remains to be seen whether the {\sc CNOT} swap itself is compatible with the postulates of black hole
complementarity  that
were by {\sc AMPS} claimed to be internally incompatible. In particular one may worry whether
the swap can be achieved by any local operation or if at least the necessary non-locality can be confined to
a region sufficiently close by the event horizon.

This paper addresses the question what has to happen in the near horizon region to
remove the firewall and to have the outgoing radiation be pure. This question can be
formulated as a requirement on the Bogoliubov transformations, which can be read as
a boundary condition that enables an interaction.  We will see that the swap that is necessary to remove the firewall can indeed 
be created locally and in the
near horizon region, and this boundary and its effect can be interpret as the 
stretched horizon.

Alas, it also will become clear that the same operation that removes
the firewall when the state at ${\cal I}^-$ is the vacuum state, cannot also be used
to transfer infalling information into the outgoing radiation. 
It thus remains to be seen whether information can be preserved
for arbitrary initial states, including the information of the infalling
observer themselves. The aim of this paper is thus not to solve the
black hole information loss problem but to identify conditions
for the absence of the firewall.

We use units in which $m_{\rm pl} = \hbar = c = 1$.

\section{Nomenclature}

As usual we use advanced and retarded time coordinates. We denote with $(U,V)$ the coordinates inside the 
collapsing matter distribution, and with $(u,v)$ the  coordinates outside, where
\beqn
v = r^* + t \quad,\quad u = r^* - t \quad,\quad r^* = r + 2M + 2M \ln((r- 2M)/2M) \quad,
\eeqn
and $r,t$ are the standard Schwarzschild-coordinates. The surface of the collapsing matter is located at $R(u,v)=0$ (see Figure \ref{fig1}). 
A complete set of in-modes that solve the wave-equation of a massless scalar field $\Phi$ in the black hole background can be composed using
up-modes and dn-modes  (following the notation of \cite{Novikov})
\beqn
\phi_{\rm dn}(v) &\sim& \Theta(v)  e^{- {\rm i} \omega \ln (\kappa v)/\kappa} \quad, \\
\phi_{\rm up}(v) &\sim& \Theta(-v) e^{{\rm i} \omega \ln (- \kappa v)/\kappa}  \quad,
\eeqn
where $\Theta$ is the Heaviside step function, $4M = 1/\kappa$, and normalizations are ommited because we will not need them (we only
care about the difference to the standard case). The up-modes cover only the region before
the last escaping ray at $v=0$, and the dn-modes only cover the space after that ray. 

Here and in the following
we will suppress an index $\vec k$ referring to the momentum of the mode, or $\omega, l, m$ referring
to its frequency and spherical harmonics composition respectively. One thus has to keep in mind that
the following vectors and matrices are a compressed notation that represents 
infinitely many entries. We will neglect spherical harmonics and backscattering on the
potential wall.

It is common to use the linear combinations
\beqn
\phi_{\rm d}(v) &=& c \left( \phi_{\rm dn} + w \bar \phi_{\rm up} \right) \quad,
\quad \phi_{\rm p}(v) = c \left(  \phi_{\rm up} + w \bar \phi_{\rm dn} \right) \\
w &=& e^{- \pi \omega /\kappa} \quad,\quad c = (1- w^2)^{-1/2} \quad,
\eeqn
(a bar denotes complex conjugation) for the in-modes, which is sometimes referred to 
as Wald's base \cite{Wald:1975kc}. The usefulness of these (orthogonal) modes comes from them being of positive frequency with respect to $v$,
a consequence of the equalities
\beqn
\int_{-\infty}^\infty {\rm d}v e^{ - {\rm i} \omega v} \phi_{\rm d}(v) = \int_{-\infty}^\infty {\rm d}v e^{ -{\rm i} \omega v} \phi_{\rm p}(v) = 0\quad.
\eeqn
In these integrals, the prefactor $w$ eats up the branch-cut
contribution of the logarithm, resulting in a cancellation of the two parts of the integrals over $v>0$ and
$v<0$. For details, see \cite{Novikov}, Appendix H.4. Since $\phi_{\rm d}$ and $\phi_{\rm p}$
are of positive frequency, one can use them to define the in-vacuum while at the same time simplifying the
Bogoliubov-transformations. The field can be expressed in this basis as
\beqn
{\Phi}_{\rm in}^{\rm T} =  (\phi_{\rm d}, \phi_{\rm p}) \quad,
\eeqn
where T denotes transposition and is just used for better display of the vector. 
The corresponding annihilation and creation operators will be denoted as 
\beqn
{\bf a}_{\rm in} = (a_{\rm d}, a_{\rm p}) \quad,\quad {\bf a}^\dag_{\rm in} = (a^\dag_{\rm d}, a^\dag_{\rm p}) \quad
\eeqn
and are by default row-vectors.
For the out-base one can use
\beqn
\Phi^{\rm T}_{\rm out} = (\phi_{\rm out}, \phi_{\rm dn}) \quad,
\eeqn
where $\phi_{\rm out} \sim e^{- {\rm i} \omega u}$.
The in-base and out-base are related by the Bogoliubov transformation
\beqn
\Phi_{\rm in} = {\bf \bar A}^{\rm T} \Phi_{\rm out} - {\bf B}^{\rm T} \bar \Phi_{\rm out} \quad. \label{Bogn}
\eeqn 
Without backscattering, these matrices can be calculated from the scalar products of the in-modes and
out-modes and are \cite{Novikov}
\beqn
{\bf A} = c
\left(
\begin{array}{cc}
0 & 1  \\
1 & 0
\end{array}
\right) \quad, \quad {\bf B} = 
-wc \left(
\begin{array}{cc}
1 & 0 \\
 0 & 1
\end{array}
\right) \quad.
\eeqn
The important physical combinations of the Bogoliubov matrices are firstly
\beqn
{\bf B} \bar {\bf B}^{\rm T} = c^2 w^2 \left(
\begin{array}{cc}
1 & 0 \\
0 & 1
\end{array}
\right) \quad.
\eeqn
The out-out component of ${\bf B} {\bf \bar B}^{\rm T}$ gives the particle-spectrum of the in-vacuum
for the out-observer
\beqn
\langle 0_{\rm in} | a^\dag_{\rm out} a_{\rm out} | 0_{\rm in} \rangle = \left({\bf B} {\bf \bar B}^{\rm T} \right )_{\rm{out,out}} = 
c^2 w^2 = \frac{1}{\exp( 2 \pi \omega /\kappa) - 1}\quad. 
\eeqn
Here $|0_{\rm in} \rangle$ denotes the vacuum annihilated by ${\bf a}^\dag_{\rm in}$. One finds the expected
 thermal spectrum with temperature $T=\kappa/(2 \pi)$.

And secondly, the  $S$-matrix can be written as  \cite{Novikov}
\beqn
|0_{\rm in} \rangle = S | 0_{\rm out} \rangle = e^{{\rm i} {\bf W}} e^{\bf F}
\exp\left( \frac{1}{2} {\bf a}_{\rm out}^\dag {\bf V} ({\bf a}_{\rm out}^\dag)^{\rm T} \right) | 0_{\rm out} \rangle  \quad,\quad
{\bf V} = - {\bf  \bar B} \left( {\bf A}^{-1} \right)~. \label{Smatrix}
\eeqn
Here ${\bf W}$ is a pure phase and not relevant in the following, and
\beqn
{\bf F} = \frac{1}{2} {\bf a}_{\rm out} {\bf A}^{-1} {\bf B} {\bf a}^{\rm T}_{\rm out} + {\bf a}_{\rm out}^\dag \left( ({\bf A}^{-1})^{\rm T} - {\bf 1} \right) {\bf a}_{\rm out}^{\rm T} 
\eeqn
will not contribute to the transition element after normal ordering.
With the above
it is
\beqn
 {\bf A}^{-1} 
=  \frac{1}{c}\left(
\begin{array}{cc}
 0 & 1 \\
1 & 0
\end{array}
\right) \quad,
\eeqn
and thus
\beqn
{\bf V} 
= w \left( 
\begin{array}{cc}
0 & 1 \\
1 & 0 
\end{array}
\right) \quad.
\eeqn
The relevant property that is eventually responsible for the outgoing state being mixed is that ${\bf V}$ is non-diagonal. Its
non-vanishing entries are in the out-dn and dn-out components. This means that the out-vacuum contains entangled pairs of
particles in the out-modes and dn-modes. If the dn-modes behind the horizon are lost to the outside observer, the state at
${\cal I}^+$ is mixed and lacks information.

\section{Requirements}

Whatever happens in the near horizon region, if unitarity is to be maintained, it must induce a suitable Bogoliubov
trafo from the usual out-state into a new out-state
\beqn
\tilde \Phi_{\rm out} = {\bf C} \Phi_{\rm out} + {\bf D} \bar \Phi_{\rm out} \label{Bog}\quad.
\eeqn
Our aim is now to identify exactly what properties this transformation must have.
\begin{enumerate}
\item The operation we are looking
for should change the correlations in the radiation received at ${\cal I}^+$, but not the spectrum. If 
it would alter the spectrum, one would run into the same problem as pointed out by {\sc AMPS},
namely that, when tracing the modes from ${\cal I}^+$ to the vicinity of the horizon, they will significantly
deviate from `empty' space (on scales below the curvature radius) and the local observer will notice. 

The reason for this problem goes back to the divergence of the stress-energy, which is the relevant
observable.
As has been shown in \cite{Candelas:1980zt,Lowe:2013zxa}, the renormalized vacuum expectation
value of the stress-energy tensor in the near horizon region is finite for the local observer if and
only if the spectrum at ${\cal I}^+$ is the normal thermal spectrum with temperature determined by
the black hole's mass. Any deviations from this thermal spectrum, regardless how small,
will be blueshifted and generate drama for an observer who has to pass through. The 
stress-energy is determined by the spectral distribution of the modes and the absence of drama
requires that this distribution should not be altered. This
means that the operation we are looking for should not create additional particles, ie it should not mix
positive and negative frequencies of the outgoing radiation, and thus ${\bf D} =0$, and
${\bf C}^{-1} = \bar{\bf C}^{\rm T}$.  The transformation (\ref{Bog}) will then 
necessarily mix positive and negative frequencies of the ingoing state.

\item The transformation (\ref{Bog}), when combined with the normal transformation (\ref{Bogn}), should
no longer result in a mixing of dn-modes and out-modes. With $\widetilde {\bf A} := {\bf C} {\bf A}$ and $\widetilde {\bf B} := {\bf C} {\bf B}$
we denote the new transformation from the in-states to the out-states as
\beqn
\Phi_{\rm in} = \bar{\widetilde{\bf A}}^{\rm T} \tilde \Phi_{\rm out} - \widetilde {\bf B}^{\rm T} \bar {\tilde \Phi}_{\rm out} \quad.
\eeqn
The second requirement then takes the form that
\beqn
\widetilde {\bf V} := - \bar {\widetilde {\bf B}} \widetilde{\bf A}^{-1} = {\rm diag} 
\eeqn
should not have off-diagonal entries. Since we do not produce additional particles and the
spectrum outside should not change, this means 
\beqn
\widetilde {\bf V} := {\bf V} \left( 
\begin{array}{cc}
0 & 1 \\
1 & 0 
\end{array}
\right) \quad.
\eeqn

\item Momentum conservation must be maintained, ie the matrix ${\bf C}$ should be proportional
to $\delta({\bf p} -  \tilde {\bf p})$, where ${\bf p}$ is the total momentum ingoing to the region close
by the horizon, and ${\bf p}$ is the total momentum outgoing from that region. Since we are using
spherical coordinates and spherical harmonics, this $\delta$-function should be read as being
factorized into spherical coordinates as well so it matches the modes.

\end{enumerate}

\section{Boundary condition}

\begin{figure}[ht]
\includegraphics[width=6.5cm]{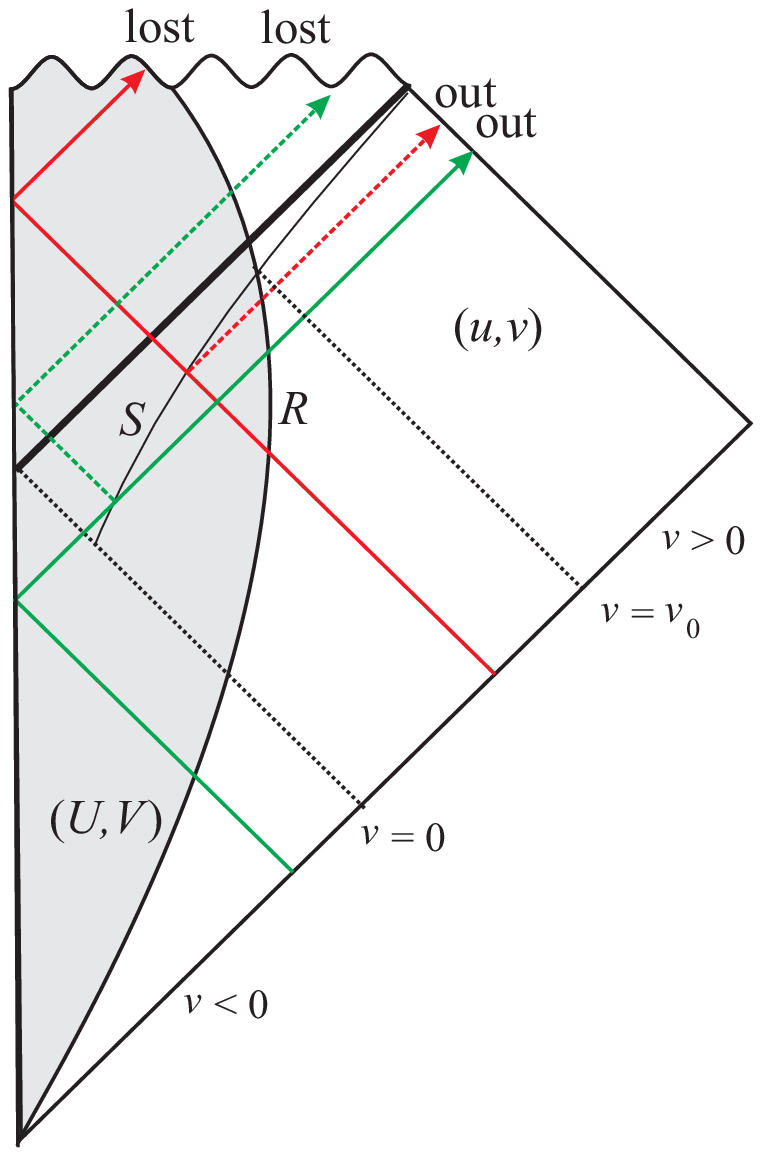} \hspace*{0.5cm} \includegraphics[width=6.5cm]{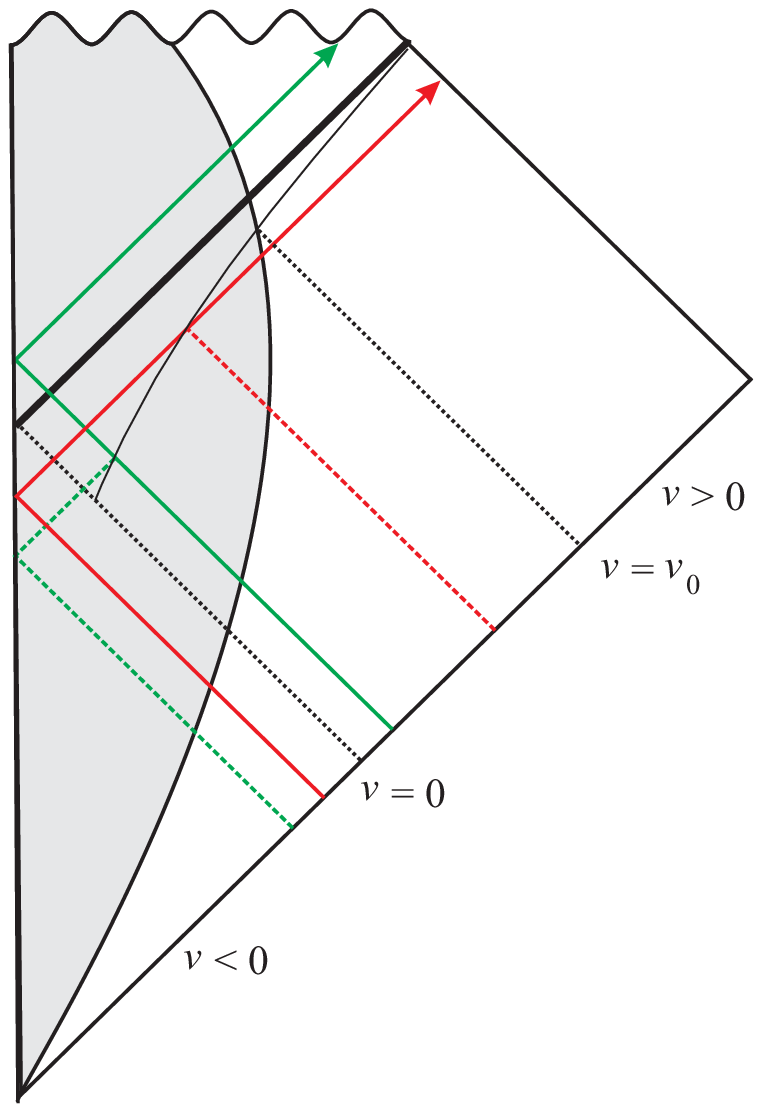}

\caption{Causal Diagram of black hole formation.  The wavy line demarks the singularity, the fat black line the
horizon, and the thin black line marked with $S$ is the stretched horizon. The grey shaded region is the inside of the
collapsing matter; its surface is at $R(u,v)=0$. The (red and green) diagonal lines depict modes. The solid parts of
these modes depict the usual ray-tracing, the dashed parts depict the new contribution from the boundary condition
at $S$. 
Left: The outgoing radiation contains information both from the
region $v \leq 0$ and $v\geq 0$. Right: The same setting but traced backwards. \label{fig1}}
\end{figure}

We will now see if there is a local boundary condition that can induce the desired 
transformation. For this, it is helpful to first depict what the transformation must
do, see Figure \ref{fig1}. For the observer at ${\cal I}^+$ to obtain all the information from ${\cal I}^-$,
an out-mode must contain information both from the region $v<0$ as well
as from the region $v>0$. This means that the stretched horizon must act
similar to a partially reflecting mirror, which is most easily seen when traced backwards,
see Figure \ref{fig1}, right. 

To obtain a boundary condition, we now put a surface at $S(u,v)=0$ and denote by $u_0, v_0$ its crossing point with the
surface of the collapsing matter 
\beqn
(u_0, v_0) : S(u_0, v_0) = 0 ~\wedge~ R(u_0,v_0) =0 \quad.
\eeqn
Outside of the collapsing matter, we locate the surface at
\beqn
S(u,v) = u - \kappa \ln(v/\kappa) + y \quad, \label{suv} 
\eeqn
where $y$ is a (real positive) constant that will only contribute a phase.
Inside, for $v_0\geq v >0$, we situate $S(U,V)=0$ at
\beqn
S(U,V) = UV - \kappa^2 e^y \quad.
\eeqn
At that surface $S=0$ we put a partial reflecting condition with the following property expressed in the out-basis as
\beqn
\Phi_{\rm out} \Big|_{\rm to} -  {\bf C} \Phi_{\rm out}  \Big|_{\rm away} = 0 \quad, \quad  \label{bdry}
\eeqn
where `to' means incoming to $S(u,v) =0$ (ie coming from ${\cal I}^-$) and `away' means outgoing from $S(u,v) =0$ (ie towards
${\cal I}^+$ and the singularity) in Figure \ref{fig1}. To connect this to our previous notation, the state
incoming to the boundary is $\tilde \Phi_{\rm out}$.

We take ${\bf C}$ to be
\beqn
{\bf C} &=&  \left( 
\begin{array}{cc}
{\rm i}/\sqrt{2} & -{\rm i}/\sqrt{2} \\
1/\sqrt{2} & 1/\sqrt{2}
\end{array}
\right) {\bf Y} \delta^4({\bf p} - \tilde {\bf p}) {\rm Vol} \quad, \label{Cmatrix} \\
{\bf Y} &=&  \frac{1}{2}\left( 
\begin{array}{cc}
1+e^{{\rm i} \omega y} & {\rm i} - {\rm i} e^{- {\rm i} \omega y} \\
{\rm i} - {\rm i} e^{{\rm i} \omega y} & 1+ {\rm i} e^{- {\rm i} \omega y}
\end{array}
\right) \quad.
\eeqn
The matrix ${\bf Y}$ is only there to adjust the phase factor. ${\bf p}$ is the total momentum of the $\Phi$-modes incoming to
the boundary and $\tilde {\bf p}$ is the total momentum of the $\tilde \Phi$-modes outgoing from the boundary in spherical
coordinates.

We will now go on to demonstrate that this boundary condition fulfills our requirements. Recall that we are suppressing
an index $\omega$ labelling the modes. The above ansatz for ${\bf C}$ means that we take the boundary condition to
not depend on the frequency.

To begin with, it is ${\bf C} {\bf \bar C}^{\rm T} = {\bf 1}$, so we can interpret this as the result of a scattering process that
comes about by some unknown physics located at $S=0$, and will not ruin unitarity.  The volume factor is there to take into account a suitable normalization procedure for plane
waves as usual. In the following we will look at a state (the vacuum state) that preserves the four momentum 
at the boundary and we will take this factor to be one, though we will come back to the requirement of
momentum conservation in the discussion \ref{disc}. Before we come to the discussion though, note that 
here and for the rest of this and the next section we are not talking about what happens to infalling information,
we are only considering the case in which the field is in the vacuum state at ${\cal I}^-$. 

If we make a plane-wave ansatz for $\Phi$, the
reflecting condition (\ref{suv}) will convert $\Phi_{\rm away}^{\rm T} = (\exp(- {\rm i} \omega u), 0) $ coming from ${\cal I}^+$ 
to 
\beqn
\tilde \Phi^{\rm T} = ({\rm i} e^{- {\rm i} \omega u}/\sqrt{2}, - {\rm i}e^{-{\rm i} \omega \ln(v/\kappa)}/\sqrt{2})
\eeqn moving towards $S=0$. As anticipated, the phase shift from $y$ is cancelled by ${\bf Y}$. This means
the boundary condition converts both out-modes and dn-modes  into combinations of out- and dn-modes at the
same frequency. Note that the location of $S$ is essential for this to work. Needless to say, this particular
scattering process would not preserve momentum because part of the wave is reflected. Alas, in the vacuum
state there will be another contribution from $\phi_{\rm dn}$ that provides the missing momentum. 
The requirement of momentum conservation in the chosen base means that the mixture of modes dependent
on $u$ and $v$ has to stay the same at the boundary. Since ${\bf C}$ mixes them in equal parts, this
means that the state ingoing to the boundary must also have had contributions in equal parts, ie
$\tilde \Phi^{\rm T} \sim (\phi_{\rm out}, \phi_{\rm dn})$, the relevant point being that the
prefactors in both components are identical.

For the annihilation and creation operators, the boundary condition performs the following mixing
\beqn
\tilde {\bf a}_{\rm out} &=& {\bf a}_{\rm out} \bar {\bf C}^{\rm T} \quad, \quad 
\tilde {\bf a}^\dag_{\rm out} = {\bf a}^\dag_{\rm out} {{\bf C}}^{\rm T} \quad.
\eeqn

The physical picture is the following. We take the in-modes and move
them forward to ${\cal I}^+$ or to the inside of the horizon respectively. On the way, 
we have to pass $S=0$ and have to take into account the boundary condition. By this, the
solutions of the wave-equation change. If we start
with defining the in-vacuum as the physically real vacuum, then the out-state that corresponds
to the in-vacuum is now a different state. It differs from the `normal' out-vacuum that contains
thermal radiation (without the boundary condition) by
the operation induced by ${\bf C}$. The early ingoing modes do not encounter a boundary
condition. This seems natural, because in the early phase the matter distribution is nothing
like a black hole. The dominant contribution for the Hawking-radiation comes from the
near-horizon region which is why the approximation around these rapidly oscillating modes
is excellent. Therefore it does not really matter what happens with the early ingoing modes
and we will not further specify this region.

In the new basis we have
\beqn
\widetilde {\bf B} \bar{\widetilde{\bf B}}^{\rm T} = \bar {{\bf C}} {\bf B} {\bf \bar B}^{\rm T} {\bf C}^{\rm T}  = 
 c^2 w^2 \left(
\begin{array}{cc}
1 & 0 \\
0 & 1
\end{array}
\right) \quad,
\eeqn
ie the spectrum remains unchanged as expected. And further
\beqn
|0_{\rm in} \rangle &=& 
e^{{\rm i} {\bf W}}
\exp\left( \frac{1}{2} {\bf a}_{\rm out}^\dag \widetilde {\bf V} ({\bf a}_{\rm out}^\dag)^{\rm T} \right)  |  0_{\rm out} \rangle \quad,\quad \label{tildes}
\eeqn
(since the boundary condition does not mix positive with negative frequencies, the ${\bf F}$ matrix does not come in) where now 
\beqn
\widetilde {\bf V} = {\bf\bar {C}} {\bf V} {\bf {C}}^{\rm T} =  w \left( 
\begin{array}{cc}
1 & 0 \\
0 & 1
\end{array}
\right) \quad.
\eeqn
Thus, with the additional boundary condition, we get pairs outside and pairs inside while the spectrum remains the same.
Inserted into the exponent of the scattering matrix (\ref{Smatrix}), the state factorizes into a
part inside and a part outside the horizon.

Inside the matter distribution, the order of operations is changed, in that a mode coming from ${\cal I}^-$ first
encounters the stretched horizon and then the surface of the matter where the usual Bogoliubov transformation
is caused. The location of the boundary is easily adjusted (again taking into account the phase shift) so as to have the same effect as outside the matter.
Again, note that the location of the boundary is relevant for the mixing to happen between the out-modes and dn-modes. The
boundary condition cannot be assigned to an arbitrary timelike surface\footnote{At least not as long as ${\bf C}$ is assumed to be independent of 
position and frequency.}.

To understand better what is going on, we can ask what in-state would be needed to get the disentangled out-state directly, 
without the boundary condition. We will call this vacuum $|\tilde 0_{\rm in} \rangle$. There are two ways to look at this. 
First, we have
\beqn
|\tilde 0_{\rm in} \rangle =  e^{{\rm i} {\bf W}}
\exp\left( \frac{1}{2} {\bf a}_{\rm out}^\dag \widetilde {\bf V} ({\bf a}_{\rm out}^\dag)^{\rm T} \right) | 0_{\rm out} \rangle \quad,
\eeqn
which, combined with the relation for the new out-vacuum (\ref{tildes}) gives
\beqn
|\tilde 0_{\rm in} \rangle =  e^{{\rm i} {\bf W}}
\exp\left( \frac{1}{2} {\bf a}_{\rm out}^\dag \left( {\bf V}-\tilde{\bf V}\right) ({\bf a}_{\rm out}^\dag)^{\rm T} \right) | 0_{\rm in} \rangle \quad,
\eeqn
where
\beqn
{\bf a}_{\rm out}^\dag \left( {\bf V}-\tilde{\bf V}\right) ( {\bf a}_{\rm out}^\dag)^{\rm T} = w \left( a^\dag_{\rm out}a^\dag_{\rm dn}  +
a^\dag_{\rm dn} a^\dag_{\rm out} -  a^\dag_{\rm out} a^\dag_{\rm out} - a^\dag_{\rm dn} a^\dag_{\rm dn} \right) \quad. \label{wtf}
\eeqn
This just means that we are adding the entangled Hawking pairs and then subtracting the disentangled ones. This expression
is however hard to interpret because it contains the out-operators. 

Another way to look at this is to first 
define the new in-state that is the normal tracing back of the new
out state
\beqn
\tilde \Phi_{\rm in} = \bar {\bf A}^{\rm T} \tilde \Phi_{\rm out} - {\bf B}^{\rm T} \bar{\tilde \Phi}_{\rm out} \quad.
\eeqn
Then we insert $\tilde \Phi_{\rm out} = {\bf C} \Phi_{\rm out}$ and transform back into the normal in-basis
with the inverse transformation
\beqn
\Phi_{\rm out} = {\bf A} \Phi_{\rm in} + {\bf B} \bar \Phi_{\rm in} \quad.
\eeqn
This gives
\beqn
\tilde \Phi_{\rm in} = {\bar{\widetilde{\bf C}}}^{\rm T} \Phi_{\rm in} - \widetilde {\bf D}^{\rm T} \bar \Phi_{\rm in} \quad,
\eeqn
with
\beqn
 {\bar {\widetilde {\bf C}}}^{\rm T} &=& \bar {\bf A}^{\rm T}{\bf C} {\bf A} - {\bf B}^{\rm T} \bar {\bf C} \bar {\bf B} \quad, \quad
 \widetilde {\bf D}^{\rm T} =  \bar {\bf A}^{\rm T} {\bf C} {\bf B} - {\bf B}^{\rm T} \bar {\bf C} \bar {\bf A} \quad.
\eeqn
Working out the products, one finds
\beqn
\widetilde {\bf C} &=& 
\frac{c^2}{\sqrt{2}}
\left(
\begin{array}{cc}
1-{\rm i}w^2 & {\rm i} - w^2 \\
1+{\rm i}w^2   & -{\rm i} - w^2
\end{array}
\right) \quad, \\
\widetilde {\bf D} &=& 
\frac{c^2 w}{\sqrt{2}}
\left(
\begin{array}{cc}
1-{\rm i} & -1+{\rm i} \\
1+{\rm i}  & -1-{\rm i} 
\end{array}
\right) \quad,
\eeqn
and
\beqn
\widetilde {\bf D} {\bar{\widetilde {\bf D}}}^{\rm T} &=& 
2 c^4 w^2
\left(
\begin{array}{rr}
1 & - {\rm i}  \\
 {\rm i}   & 1
\end{array}
\right) \quad, \label{what} \\
- \bar{\widetilde {\bf D}} {{\widetilde {\bf C}}}^{-1} &=& 
\frac{w}{1+w^2}
\left(
\begin{array}{rr}
- {\rm i} & -  1  \\
- 1   & - {\rm i}
\end{array}
\right) \quad. \label{vtt} \\
\eeqn

There is no good physical interpretation for comparing these two in-vacua, but one can extract from
this expression what one could have expected already from (\ref{wtf}), namely that (when expressed in
terms of up-modes and dn-modes), the necessary in-vacuum to create a disentangled out-vacuum
without the additional boundary condition must have had correlations between the regions $v>0$ and $v<0$.
It is thus not an initial state that we would plausibly have for a collapsing matter distribution but instead
a delicately finetuned one.

Since $S$ contains a $\ln(v)$ it will not be very close to the horizon in any general way at late times. But one can use the
constant $y$ to get it to stay as closely as desired within any finite amount of time. Since the black hole
eventually evaporates, a finite time is sufficient. 

\subsection{A simplified qubit model}

We can extract the main idea in a simplified qubit version. Instead of
considering the Hawking spectrum with a continuum of modes, we will
just look at two different states, denoted $+$ and $-$. Each of them
can either leave the black hole horizon towards future infinity (corresponding to the out-mode) 
or fall in (corresponding to the dn-mode). We will denote these four
states as $|O+\rangle$, $|O-\rangle$ and $|D+ \rangle$, $|D-\rangle$.

The usual Hawking state with entanglement between the inside and
outside region then corresponds to
\beqn
\frac{1}{\sqrt{2}} \left( \frac{1}{\sqrt{2}} \left( |O+\rangle |D+\rangle + |D+\rangle |O+\rangle \right) 
+ \frac{1}{\sqrt{2}} \left( |O-\rangle |D-\rangle + |D-\rangle |O- \rangle \right) \right) \quad.
\eeqn
The scattering at $S$ (compare to Eq. (\ref{Cmatrix})) performs the following operation
\beqn
|O+\rangle \to \frac{1}{\sqrt{2}} \left( |O+\rangle + {\rm i} |D+\rangle \right)~&,&~|O-\rangle \to \frac{1}{\sqrt{2}} \left( |O-\rangle + {\rm i} |D-\rangle \right), \\ 
|D+\rangle \to \frac{1}{\sqrt{2}} \left( |O+\rangle - {\rm i} |D+\rangle \right)~&,&~|D-\rangle \to \frac{1}{\sqrt{2}} \left( |O-\rangle - {\rm i} |D-\rangle \right). 
\eeqn
That the same operation takes place for both the $+$ and $-$ states corresponds to the
frequency-independence of the boundary condition. With this operation, the qubit version of the Hawking
state goes to
\beqn
\frac{1}{2} \left( |O+\rangle|O+\rangle + |D+\rangle |D+\rangle + |O-\rangle |O-\rangle + |D-\rangle |D-\rangle \right) \quad,
\eeqn
so the entanglement between the inside and outside modes is gone. Note that the probability to
measure any term in this expression is reduced by a factor $1/2$ over the usual case. But since the outside particles
now always come in pairs, the spectrum (number density as well as spectral distribution) remain
the same.

\section{Revisiting the firewall argument}

The firewall argument \cite{Almheiri:2012rt} in brief goes like this. If the state at ${\cal I}^+$ is a pure state
with sufficient correlations between the early and the late modes, then one can
construct an excitation in the late state and trace it back to the horizon, where it becomes
a dramatically large deviation from the local observer's vacuum. The infalling observer
must notice, thus violating the equivalence principle. The essence of this firewall problem had
been identified already in \cite{Giddings:1994pj} and was discussed previously
in \cite{Mathur:2009hf}. The important contribution from \cite{Almheiri:2012rt} is that
the explicit construction of the traced-back excitation allows to identify exactly what 
property of the state at ${\cal I}^+$ is problematic
and, seeing what the problem is, to circumvent it.

In the case with the additional boundary condition proposed here, the state at ${\cal I}^+$ does have the energy
spectrum of the Hawking radiation, and there are no correlations between early and
late modes, instead there are correlations between energy modes. This remains so also when constructing localized wave-packets;
these then always have to come in pairs at the same time, up to the uncertainty that comes from the width of the now localized
wave-packages. One can thus turn the {\sc AMPS} argument around and say that if we trace back
a late state of this disentangled vacuum, then the state close by the horizon must also
have the same energy spectrum as normally, otherwise the state at ${\cal I}^+$ must
have been different. The infalling observer cannot notice anything unusual because the
boundary condition does not mix positive with negative frequency modes and thus he
does not notice any change to the vacuum. The renormalized
stress-energy delivers the usual result because there are no excitations over the thermal
spectrum. 

The observer himself 
does not notice the crossing of the boundary because any partial
reflection of a state that is not balanced from the other side of the boundary would
violate momentum conservation. Thus, the observer, or the modes he is composed of
respectively, cannot scatter on the boundary. There is no drama, because drama
would require energy or momentum, neither of which is available. Instead of
speaking of a boundary condition which suggests it acts on all modes, it might
be more useful to think of the requirement on the stretched horizon as
facilitating an interaction that necessarily maintains momentum conservation.

Another way to look at this is that a local observer who hovers at a constant radius
 $r_0$ outside the black hole horizon uses his local modes which are blue-shifted
relative to the modes at asymptotic infinity. These modes are not complete, but they
can be completed, resulting in the same set of modes and transformations as previously. The hovering observer
thus sees the same thermal spectrum
as the observer at ${\cal I}^+$, but with temperature $T= \gamma \kappa/(2 \pi)$,
where $\gamma = (1-2M/r_0)^{-1/2}$. This 
temperature diverges as the observer gets close to the horizon, because hovering
there would require an infinite accelleration. In the normal case (without the
boundary) he cannot locally distinguish between hovering nearby the black hole
horizon and acceleration in flat space. He sees the same as the Rindler observer
in Minkowski space, and the equivalence principle holds. 

The boundary condition does the same to the local observer's modes as it does
to the Hawking modes: Only suitable combinations of out-modes and dn-modes
that can scatter on the boundary, and this scattering serves to disentangle them.
The observer now finds himself in a thermal bath, but one that has pairwise
correlations. Can he measure the difference? 

First we note that the equivalence principle in its usual form 
is ambiguous on exactly what the local observer should see. He
should see the same as in flat space, but flat space with a boundary condition
(think of the moving mirror or the Casimir effect) is still flat space, yet it
makes a difference to the observables. If the local observer would come
to conclude that he is in Rindler space with a (semi-transparent) boundary
condition, would or wouldn't this amount to a violation of the equivalence
principle? 

Having said that, let us ask though what the local observer has to do to
notice a difference to empty Minkowski-space, ie absent any boundary
conditions. 

Let us assume the observer has measured a typical particle
with energy $E \sim T$ to precision $\Delta E$. Absent the boundary condition, the next particle
is entirely uncorrelated and arrives with a probability distributed according
to the Boltzmann statistic, emitted according to the Stefan-Boltzmann law. 
On the average, it takes a time of $\Delta t \sim 1/(M^2 T^3)  \sim 16 M$ until
the black hole emits a particle of energy $T$, where $t$ is the coordinate time, 
not the observer's proper time, which is $\tau = t \gamma$. The observer at a radius $r_0$
obtains the emitted paticles at a ratio reduced by a factor $(2 M)^2/r_0^2$ relative to the
flux nearby the horizon. The farther away he is from
the horizon, the longer he has to wait for the next particle since the apparent
luminosity decreases. 

The next particle
thus comes on the average within a time window $\Delta t \sim 16 M (r_0^2/(2 M)^2)$. In the
presence of the boundary condtion, it has (with probability 1/2) the same
energy as the previous one, up to the uncertainty $\Delta E$.  If the uncertainty $\Delta E$
is comparable to $T$, the observer cannot tell the difference to the uncorrelated case (or the
statistical significance of his conclusion is very low, to be more precise). The observer
can thus not make the detector arbitrarily small. It has to have at least have a size of $\sim 16 M$, where the size is measured in the asymptotically flat 
coordinates. Taken together this means the local observer's measurement needs at least
extensions $\Delta r \sim 16 M$ and $\Delta t  \sim 16 M (r_0^2/(2 M)^2)$.

We now have to compare this to the typical size of the locally flat region, the region
within which curvature effects are negligible. To this end, we can estimate the tital
forces acting on the local observer's apparatus, caused by the difference in acceleration between
the two radial positions at distance $\Delta r$. If the observer was accelerating in flat space, the same
acceleration for both radial positions $r_0$ and $r_0 + \Delta r$ would not create any relative motion. In the
Schwarzschild background, the same acceleration between both radial positions will slowly stretch the apparatus.
This is essentially geodesic deviation, except that the apparatus is not moving on
a geodesis. 

In the limit of small velocities, the difference in acceleration is approximately given by
\beqn
\frac{d^2 (\Delta r)}{dt^2} \sim \frac{M}{r_0^2} - \frac{M}{(r_0 + \Delta r)^2} \quad,
\eeqn
and thus over a time of $\Delta t$, the stretch that the detector aquires is
\beqn
\Delta r (\Delta t) \sim \left( \frac{M}{r_0^2} - \frac{M}{(r_0 + \Delta r)^2} \right) \Delta t^2 \quad.
\eeqn
Since $r_0 > 2M$, one finds  that $\Delta r (\Delta t)/\Delta r > 1$. This means that the observer
can measure the effects of curvature on his detector in the time he has to wait for sufficiently
many particles.

So the local observers still see a thermal spectrum with the redshifted temperature $T= \gamma \kappa/(2 \pi)$.  They
can distinguish the entangled from the disentangled
thermal spectrum, but cannot do it locally.

\section{Discussion} 
\label{disc}

So we have been able to construct a local boundary condition at a time-like surface
close by the black hole horizon that reconciles the purity of the outgoing radiation
with that of the ingoing vacuum state, thus removing the firewall problem. 

This
became possible by noting that a well-behaved stress-energy requires a thermal
energy spectrum \cite{Candelas:1980zt,Lowe:2013zxa}, but not necessarily a thermal state. While
a thermal spectrum is normally composed of uncorrelated particles, there are
correlated states that have the same energy spectrum. The spectral energy distribution
just does not completely specify the state. To see how this can be consider a
sequence of uncorrelated (fair) coin tosses. Now consider a sequence 
of tosses with always twice the same result in a row. The probability distribution (the spectrum) in both cases is 
the same. The difference is in the correlation between subsequent tosses (the entanglement). To remove
the firewall we need an out-state that has correlations but still a thermal energy spectrum. The
state considered by {\sc AMPS} is not of this type because it necessarily has excitations over
thermality at late times.

The construction
suggested here represents an explicit example for the general argument in  \cite{Hossenfelder:2012mr}
that the problematic assumption that leads to the existence of a firewall is not to be
found in the axioms of black hole complementarity but in the specific way that
the state at ${\cal I}^-$ was assumed to be entangled.

However, a series of pure states with a thermal spectral distribution might
be pure but it still does not contain information (other than the temperature). That is
to say, disentangling the vacuum by a suitable boundary condition on the stretched horizon
may remove the firewall but in and by itself that does not address the black hole information
problem. For this, one would have to find a way to actually encode information from 
infalling matter (observers) into the outgoing state while still keeping the radiation pure
and the spectrum thermal. 

This brings us back to the question what happens to in-states other than the vacuum. As noted
previously, due to momentum conservation, nothing will happen
to the state itself at the stretched horizon. There cannot be anything happening, because
there is no energy there that could make anything happen. However, recall also that
we assumed the ${\bf C}$ matrix does not depend on the frequency. This does not necessarily have to be so. One could for example add a relative
phase between different modes. This would not affect the spectrum but still
constitute an observable change. 

If we decide to either shift or not shift a certain mode by, say, a phase of ${\rm i} \pi$, then
we have one bit of information per mode that can be distinguished and measured (each mode
is either shifted or not relative to one reference mode). The
black hole of mass $M$ typically emits $M^2$ particles with approximate energy of $1/M$, which
gives about $2^{M^2}$ states. This estimate shouldn't be taken too seriously though because
we have neither taken into account exactly how good a phase shift can be measured given
the uncertainty of wave-packets that can be constructed, nor have we accounted for the
spherical harmonics. This is just to illustrate that it seems possible to use the freedom
in ${\bf C}$ to construct sufficiently many different states. In the end though, the
boundary condition must come about by some quantum property of the background
geometry, and the number of states must be determined by the possible microstates
of the background.

Of course this only moves the problem elsewhere, because then we
have to wonder how the information got from the in-state into the
boundary condition. It is beyond the scope of the paper to devise a mechanism
for how the stretched horizon may read and release information of matter
passing through, but we can take away that, to resolve the black hole information
loss problem, the boundary condition must
depend on the ingoing state.

An entirely different question is whether there is an equation of motion or an interaction term that generates the boundary condition (\ref{bdry})? 
There
are some examples of different reflecting conditions in \cite{Haro:2007ue,Haro:2008zza} but these cannot
be directly applied here. One can come up with an equation by noting that when we reverse the boundary condition, we must get a
solution of the normal wave-equation. Thus
\beqn
\partial_u \partial_v \Phi_{\rm in}\left({\bf 1} \Theta(S) + {\bf C}^{-1} \Theta(-S) \right) = 0 \quad.
\eeqn  
One can then take the derivatives, but this does not shed much light on the microscopic
origin of the ${\bf C}$-operation

The here proposed boundary condition also circumvents another
difficulty with the setting proposed in \cite{Verlinde:2013uja}, that is the doubling
of the initial state. The problem with this is that in a space-time with several
black holes each of them would swallow a copy of the initial state and it is not
clear how to generalize the construction of the Hilbert-space then. Here, we
instead split the incoming states and recombine them
differently so as to disentangle the Hawking-radiation, which makes the
doubling unnecessary.

Finally, it would be interesting to see whether the boundary condition
proposed here can be found by means of the gauge-gravity duality or
is related to the recent development reported in \cite{Papadodimas:2013wnh,Papadodimas:2013jku}.

\section{Conclusion}

The requirement that
the state at ${\cal I}^+$ which corresponds to the in-vacuum is a disentangled, pure, version of the usual mixed Hawking state with the same
spectral distribution allows to formulate a boundary condition in the region
close by the horizon. This boundary enables an interaction and prevents a mismatch between the traced-back pure
radiation at ${\cal I}^+$ and the moved-forward vacuum from ${\cal I}^-$ that gives rise to the
firewall paradox; it can be interpreted as the stretched horizon. While this in an by itself does not address the black hole information loss
problem, the necessary boundary condition is not unique, and this non-uniqueness
might potentially be used to encode information in the outgoing radiation.

\section*{Acknowledgements}

I thank Raffael Bousso, Steve Giddings, Steve Hsu, Ted Jacobson and Don Marolf for feedback, and Joe Polchinski
for spotting several mistakes in an early draft. I thank Stefan Scherer and L\'arus Thorlacius for helpful discussions.

\end{document}